\documentclass[11pt,a4paper]{article}
\usepackage[dvips]{graphicx}
\usepackage[small]{caption}

\newcommand{\hdmo}{HEIDELBERG-MOSCOW experiment~}

\def \onbb {$0\nu\beta\beta$ }

\def \beq {\begin{equation}}
\def \eeq {\end{equation}}

\def \gs  {Gran Sasso Underground Laboratory}

\begin{document}

\title{Measurement of the $^{214}\rm{Bi}$ spectrum in the energy region
  around the Q-value of $^{76}\rm{Ge}$ neutrinoless double-beta decay.}

\author{H.V. Klapdor-Kleingrothaus
\footnote{Spokesman of HEIDELBERG-MOSCOW and GENIUS Collaborations, 
E-mail: klapdor@gustav.mpi-hd.mpg.de,
Home-page: $http://www.mpi-hd.mpg.de.non\_acc/$} 
, O. Chkvorez, I.V. Krivosheina, C. Tomei \\
 \emph{Max-Planck-Institut f\"ur Kernphysik, PO 10 39 80,}\\
  \emph{D-69029 Heidelberg, Germany}}

\date{18.05.2003}

\maketitle


\begin{center}
\section*{Abstract}
\end{center}
	In this work we present the results obtained measuring the
	$^{214}\rm{Bi}$ spectrum from a $^{226}\rm{Ra}$ source with a high
	purity germanium detector. Our attention was mostly focused on the 
	energy region around the Q-value of
	$^{76}\rm{Ge}$ neutrinoless double-beta decay (2039.006\,keV). The
	results of this measurement are strongly related to the first
	indication for the neutrinoless double beta decay of $^{76}\rm{Ge}$,
	given by a recent analysis 
\cite{Evidence,KK02-PN,KK02-Found,KK-BigArt02} 
	of the data collected
	during ten years of measurements from 
	the HEIDELBERG-MOSCOW experiment.


\section{Introduction}


	A recent analysis 
\cite{Evidence,KK02-PN,KK02-Found,KK-BigArt02} 
	of the data collected during ten
	years of measurements by the HEIDELBERG-MOSCOW experiment, at the
\gs, 
	yields a first indication for the neutrinoless double beta
	decay of $^{76}\rm{Ge}$. An important point of this analysis is
	the interpretation of the background, in the region around the
	Q-value of the double beta decay (2039.006(50)\,keV, see 
\cite{Qvalue}), 
	as containing several weak photopeaks. It was suggested in
\cite{Evidence}, 
	and has been shown in 
\cite{KK02-PN,KK02-Found,KK-BigArt02}, 
that four of these
	peaks are produced by a contamination from the isotope
	$^{214}\rm{Bi}$, whose lines are present throughout the
	HEIDELBERG-MOSCOW background spectrum. 

	In a following paper
\cite{26people}, 
	some criticisms to 
\cite{Evidence,KK02-PN,KK02-Found,KK-BigArt02} 
	were raised.
	One of the main points mentioned was that the intensities 
	of the weak bismuth lines between 2000 and 2080\,keV, 
	as deduced from the intensities of the strong lines 
	and the branching ratios from the Table of Isotopes 
\cite{TOI}, 
	were much lower than observed in
	the HEIDELBERG-MOSCOW background spectrum. It was shown, in a
	reply to these criticisms 
\cite{replay}, 
	that the simple estimate of 
\cite{26people} 
	about the intensities of the weak bismuth lines
	was not correct. It did not take into account the long-known
	spectroscopic effect of the true coincidence summing (TCS, see 
\cite{TCS}) 
	and, most important, the localization of the bismuth contamination 
	inside the HEIDELBERG-MOSCOW experiment.

	With a careful
	simulation of the experimental setup, it was shown that, if the 
	contamination is localized in the copper cap of the detectors, the 
	expected values for the four peak-areas in the region of interest 
	were compatible (within two sigma) with the measured values (see 
	Table 1 of 
\cite{replay} 
	and Table 7 in 
\cite{KK02-Found,KK-BigArt02}). 

	In this work we performed a {\it measurement} 
	of a $^{226}\rm{Ra}$ source with a high-purity 
	germanium detector. 
	The aim of this work is to study the spectral shape of the lines in 
	the energy region from 2000 to 2100\,keV and, most important, to 
	show the difference in this spectral shape when changing the 
	position of the source with respect to the detector, 
	and to verify the effect of TCS for the weak $^{214}{Bi}$ 
	lines seen in the \hdmo experiment.

	The activity of the $^{226}\rm{Ra}$ source is 
	95.2\,kBq. 
	The isotope $^{226}\rm{Ra}$ appears in the 
	$^{238}\rm{U}$ natural decay chain and from its decays also 
	$^{214}\rm{Bi}$ is produced. The $\gamma$-spectrum of $^{214}\rm{Bi}$ 
	is clearly visible in the $^{226}\rm{Ra}$ measured spectrum. 
	We also performed a simulation of our measurement with the GEANT4 
	simulation tool and we find a good agreement between the 
	simulation and the measurement. 
	The results of this measurement confirm that the criticism by 
\cite{26people}
	is wrong and the analysis of 
\cite{Evidence,KK02-PN,KK02-Found,KK-BigArt02}
	of the double beta experiment is correct.


\section{The Isotope $^{214}\rm{Bi}$ and the TCS Effect}


	$^{214}\rm{Bi}$ is a naturally occurring isotope: 
	it is produced in the 
	$^{238}\rm{U}$ natural decay chain through the $\beta^-$ decay of
	$\;^{214}\rm{Pb}$ and the $\alpha$ decay of $\;^{218}\rm{At}$.
	With a subsequent $\beta^-$ reaction, $^{214}\rm{Bi}$ 
	decays then into
	$\;^{214}\rm{Po}$ (the branching ratio with respect to the $\alpha$
	decay into $^{210}\rm{Tl}$ is 99.979\%). The decay, however, does not
	lead directly to the ground state of $\;^{214}\rm{Po}$, but to its
	excited states. From the decays of those
	excited states to the ground state we obtain the well known
	$\gamma$-spectrum of $^{214}\rm{Bi}$ which contains 
	more than hundred lines.

	As one can see in Table 
\ref{tab1}, 
	in the energy region around the Q-value of the 
\onbb 
	decay (2000-2100\,keV), four $\gamma$-lines and 
	one E0 transition with energy 2016.7\,keV are expected. 
	The E0 transition 
	can produce a conversion electron or a electron-positron pair but it 
	could not contribute directly to the $\gamma$-spectrum 
	in the considered energy 
	region if the source is located outside the detector active volume. 

\begin{table}[!htp]
  \begin{center}
    \begin{tabular}{|c|c|}
\hline
      &  \\
      Energy (keV) & Intensity (\%) \\
      &  \\
      \hline
      &  \\
      2010.71 (15) & 0.050 (6)  \\
      2016.7 (3) & 0.0058 (10)  \\
      2021.8 (3) & 0.020 (6) \\
      2052.94 (15) & 0.078 (11) \\
      2089.7 (2) & 0.050 (6) \\
      &  \\
\hline
\end{tabular}
\caption[]{$\gamma$-lines from $^{214}\rm{Bi}$ in the energy
      region from 2000 to 2100\,keV around the Q-value
      of $^{76}\rm{Ge}$. The number in parenthesis is the error on the
      least significant digit (see Table of Isotopes 
\cite{TOI}).}
\label{tab1}
\end{center}
\end{table}

	The intensity of each line is defined as the number of
	emitted photons, with the corresponding energy, per 100 decays of
	the parent nuclide.

\begin{figure}[t]
\begin{center}
\includegraphics[width=11cm,height=7cm]{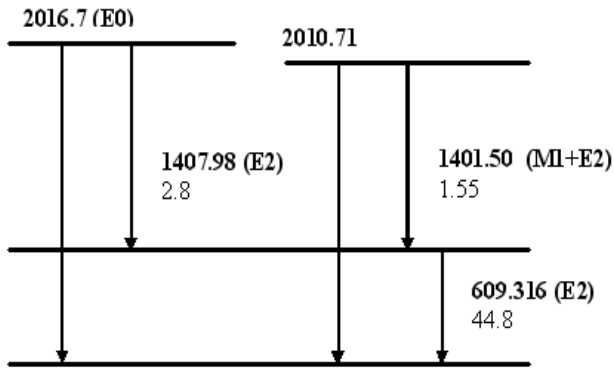}
\end{center}

\vspace{-0.5cm}
\caption[]{
	Simplified decay scheme for the two transitions 2010.7\,keV 
  	and 2016.7\,keV in $^{214}{Po}$, 
	formed by $\beta^{-}$ decay of $^{214}{Bi}$. } 
\label{levels}
\end{figure}

	When coming to the measurement, we have to
	consider the efficiency of the detector (which depends on the size
	of the detector and on the distance source-detector) and the
	effect called True Coincidence Summing (TCS). The TCS effect is
	described in detail in 
\cite{TCS}. 
	The lifetimes of the atomic
	excited levels are much shorter than the resolving time of the
	detector. If two gamma-rays are emitted in cascade, there is a
	certain probability that they will be detected together. If this
	happens, then a pulse will be recorded which represents the sum of
	the energies of the two individual photons, instead of two
	separated pulses with different energies. The TCS effect can
	result both in lower peak-intensity for full-energy peaks and in 
	bigger peak-intensity for those transitions 
	whose energy can be given by 
	the sum of two lower-energy gamma-rays. In our case, the lines at 
	2010.7\,keV and 2016.7\,keV (see Fig. 
\ref{levels}) 
	can be given by 
	the coincidence of the 609.312\,keV photon (strongest line, 
	intensity = 46.1\%) with the 1401.50\,keV photon (intensity = 
	1.27\%) or with the 1407.98\,keV photon (intensity = 2.15\%).

	The degree of TCS depends on the probability that two
	gamma-rays emitted simultaneously will be detected simultaneously.
	This is a function of the geometry and of the solid angle
	subtended at the detector by the source. For this reason, the
	intensities of the two lines mentioned above (2010.71\,keV and
	2016.7\,keV) are expected to depend on the position of the source
	with respect to the detector.


\section{The measurement}


	The $^{226}\rm{Ra}$ $\gamma$-ray spectra were measured using a
	$\gamma$-ray spectroscopy system based on an HPGe detector
	installed in the operation room of the \hdmo 
	in \gs, Italy. 
	The coaxial germanium detector has an
	external diameter of 5.2\,cm and is 4.9\,cm high. The distance
	between the top of the detector and the copper cap is 3.5\,cm. The
	relative detection efficiency of the detector is 23\% and the
	energy resolution is 3.6\,keV for the energy range 2000-2100\,keV.\\
	The electronics system consists of a linear spectroscopy amplifier
	and an ORTEC MCA board installed in a personal computer. The
	$\gamma$-ray pulses from the preamplifier are shaped into
	semi-Gaussian by the amplifier. To reduce pile-ups the shaping
	time 2 $\mu$sec was chosen. The energy threshold was set at 120
	keV to reduce the MCA dead-time. Maximum count rate and dead-time
	during the measurements were not higher then 10000\,cps and 13\%, 
	respectively.


\begin{figure}[!htp]
\begin{center}
\includegraphics[width=7cm,height=9cm]{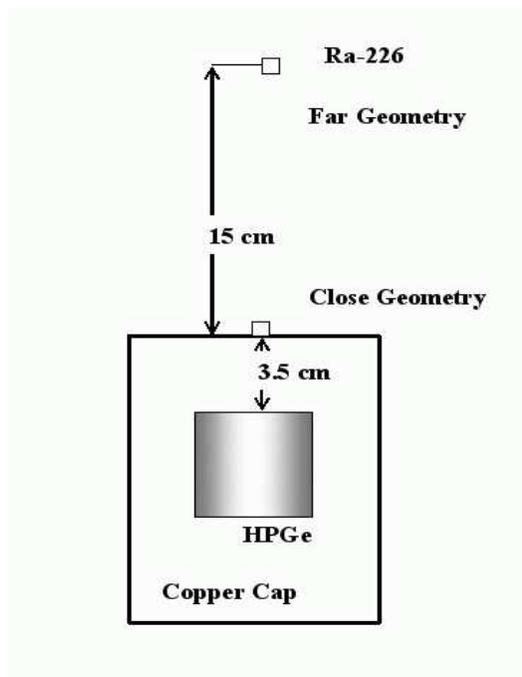}
\end{center}
\caption[]{
	Detector setup for the far geometry and the close geometry
  	measurements.} 
\label{setup}
\end{figure}


	In a first step the source was positioned on the
	top of the detector, directly in contact with the copper cap
	(close geometry). In a second step the source was moved 15\,cm away
	from the detector cap (far geometry). Fig. 
\ref{setup} 
	shows the experimental setup. 
	To collect a high statistics in the considered region, 
	the duration of the measurements were 60000\,sec and 170000\,sec 
	for close and far geometries, respectively. 
	The spectra were processed by
	the Aptec-Demo MCA Analysis Software, allowing separate overlapping
	$\gamma$-lines.


\section{Results and Discussions}

\begin{figure}[!htp]
\begin{center}
\includegraphics[width=9cm,height=7cm]{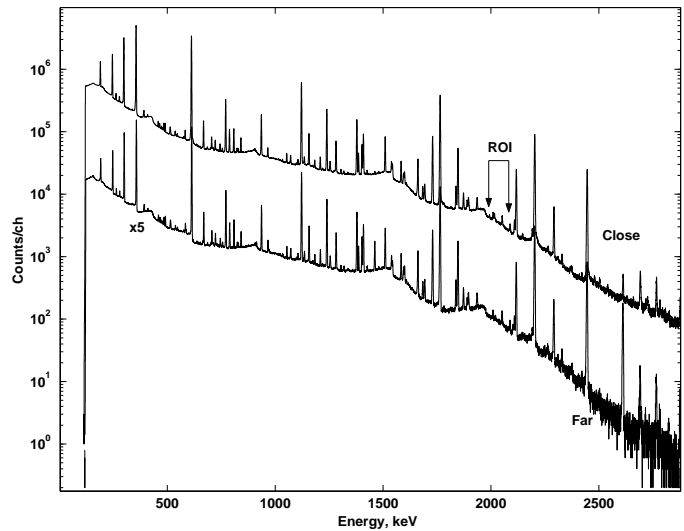}
\end{center}
\caption[]{Full measured $^{226}\rm{Ra}$ spectra. 
	The region of interest (ROI) is shown in detail in Fig. 
\ref{RaExper}.} 
\label{Full_Ra}


\end{figure}
\begin{figure}[!htp]
\begin{center}
\includegraphics[width=9cm,height=7cm]{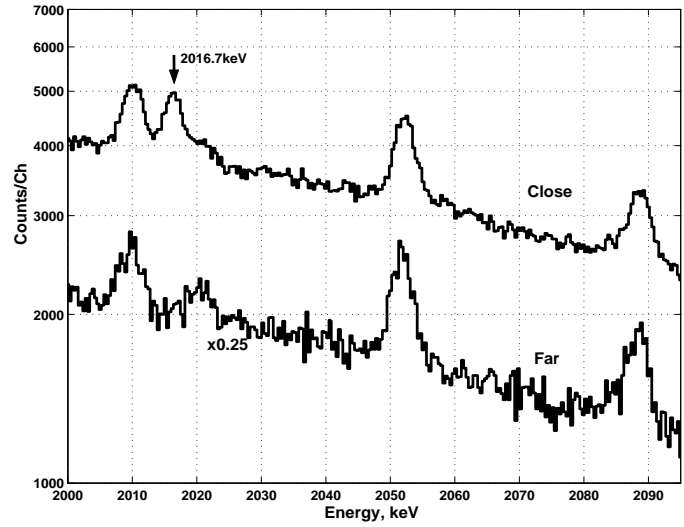}
\end{center}
\caption[]{Measured $^{226}\rm{Ra}$ spectrum in the energy range
  	from 2000 to 2100\,keV. The upper spectrum corresponds to the close
  	geometry, the bottom spectrum to the far geometry (see text).
  	The weak lines from $^{214}\rm{Bi}$ are
  	nicely visible, together with the effect of the true coincidence
  	summing at 2010.71 keV and 2016.7\,keV.} 
\label{RaExper}
\end{figure}

	The full measured spectra of the Ra source are shown in
	Fig. 
\ref{Full_Ra}, 
	the energy region of interest is shown in
	Fig. 
\ref{RaExper}. 
	One can easily appreciate the difference in the
	spectral shape when going from the far geometry (bottom spectrum) to
	the close geometry (upper spectrum). In the first case (far), 
	the line at 2016.7\,keV is completely absent 
	and the relative intensities of the other
	four lines are in perfect agreement (see Tables 
\ref{tab2} 
	and
\ref{tab3}) 
	with the numbers given in the Table of Isotopes. 
	In the second case (close), the line at 2016.7\,keV 
	is clearly visible and the relative intensities of the
	lines are globally modified (see again Tables 
\ref{tab2} and 
\ref{tab3}).


\begin{figure}[!htp]
\begin{center}
\includegraphics[width=9cm,height=7cm]{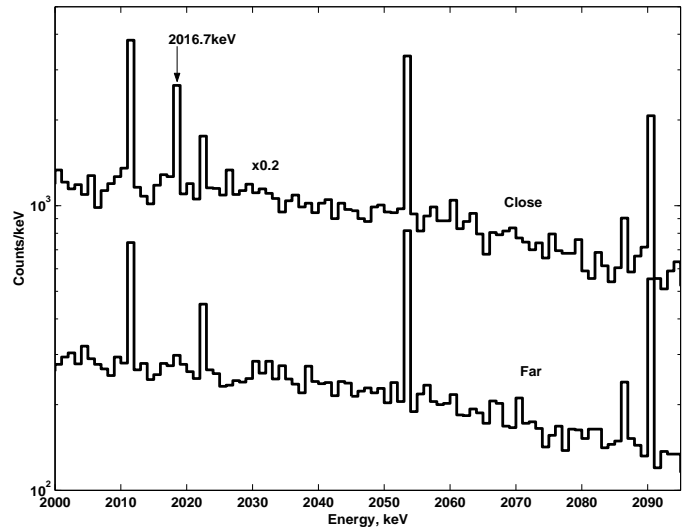}
\end{center}
\caption[]{
	Simulated $^{214}\rm{Bi}$ spectrum in the energy range
  	from 2000 to 2100\,keV. The upper spectrum corresponds to the close
  	geometry, the bottom spectrum to the far geometry (see text).} 
\label{BiSim}
\end{figure}

	A simulation with the GEANT4 Monte Carlo simulation
	tool, which reproduces the precise geometry of the experimental
	setup, has been performed and compared to the results of the
	measurements. Only the contribution of the isotope $^{214}\rm{Bi}$
	was included in the simulation. $5\times10^{7}$ $^{214}\rm{Bi}$
	decays were started for the close geometry simulation and
	$5\times10^{8}$ events were started in the case of far geometry.
	The data-file containing the complete $^{214}\rm{Bi}$
	$\gamma$-spectrum was taken from the ENSDF library and is
	available on the net 
\cite{NNDC}. 
	The results of the simulation
	are reproduced in Fig. 
\ref{BiSim}. 
	Note that the simulated
	spectrum does not take into account the resolution of the
	detector. The agreement between the simulated spectra and the
	measured ones with respect to the relative intensities of the
	lines is clear. 
	A comparison of the intensities is given in Table 
\ref{tab2} 
	for the far geometry and in Table 
\ref{tab3} 
	for the close geometry.


\begin{table}[!htp]
\begin{center}
\begin{tabular}{p{11cm}}
\hspace{0.5cm} Far Geometry (source 15\,cm away from the detector cap) \\
\end{tabular}
\begin{tabular}{c|c|c|c}
  \hline
  & & & \\
  Energy (keV) & Rel.Int.(TOI)($^{*}$) & Rel.Int.(Exp.) & Rel.Int.(Sim.)  \\
  & & & \\
  \hline
  & & &  \\
  2010.71  & 0.64  $\pm$ 0.17  & 0.68 $\pm$ 0.12 & 0.80  $\pm$ 0.09 \\
  2016.7   & 0.074 $\pm$ 0.023($^{**}$) & $<$ 0.03 & 0.06 $\pm$ 0.04 \\
  2021.8   & 0.26  $\pm$ 0.11  & 0.31 $\pm$ 0.09 & 0.31  $\pm$ 0.06 \\
  2052.94  & 1                 & 1               & 1                \\
  2089.7   & 0.64  $\pm$ 0.17  & 0.61 $\pm$ 0.14 & 0.66 $\pm$ 0.08  \\
  & & & \\
  \hline
\end{tabular}
\caption[]{
	The relative intensities of the weak $\gamma$-lines from
  	$^{214}\rm{Bi}$ in the energy region from 2000 to 2100\,keV are
  	calculated with respect to the line at 2052.94\,keV. A comparison
  	between the expectations from the Table of Isotopes, the measurement
  	and the simulation is made.

  	($^{*}$) Note that the intensities from the TOI 
  	can be compared with the measured intensities only in the case of
  	the far geometry, where the TCS effect is negligible. 

  	($^{**}$) The 2016.7\,keV transition is not a gamma-line but an E0
  	transition (conversion electron or pair). No counts from this direct
  	transition could be observed in our experimental spectrum, due to
  	the short range of the electrons. For this reason a measured value
  	compatible with zero, for the far geometry, 
	is in perfect agreement with
  	expectations.}
\label{tab2}
\end{center}
\end{table}

	The intensities of the five lines in the region 2000-2100\,keV 
	have been
	normalized to the intensity of the line at 2052.94 keV.
	In the case of the far geometry measurement (see Table 
\ref{tab2}),
	there is a good agreement between the relative intensities 
	from the Table of
	Isotopes and the measured (and simulated) intensities. This was
	expected because in this case the summing effect is negligible.

	In the case of the close geometry, the two lines at 2010.71\,keV 
	and 2016.7\,keV show an increase in the relative intensity 
	(by a factor of 1.5 and 10.6 respectively) 
	which is obtained only by putting the 
	source close to the detector (in this case 3.4\,cm instead of
	18.4\,cm). At this distance the summing effect plays a strong role.

	The intensities of the other lines, as shown in Table 
\ref{tab3}, 
	are still in agreement with the Table of Isotopes.
	According to the decay scheme of $^{214}\rm{Bi}$, 
	they were not supposed to
	be affected by the true coincidence summing.


\begin{table}[!htp]
\begin{center}
\begin{tabular}{p{11cm}}
\hspace{0.5cm} Close Geometry (source on the top of the detector cap) \\
\hline
\end{tabular}
\begin{tabular}{c|c|c|c}
  & & & \\
  Energy (keV) & Rel.Int.(TOI) ($^{*}$) & Rel.Int.(Exp.) & Rel.Int.(Sim.)  \\
  & & & \\
  \hline
  & & &  \\
  2010.71 & 0.64  $\pm$ 0.17  & 0.97 $\pm$ 0.07 & 1.11 $\pm$ 0.13 \\
  2016.7  & 0.074 $\pm$ 0.023  ($^{**}$) & 0.79 $\pm$ 0.08 & 0.63 $\pm$ 0.09 \\
  2021.8  & 0.26  $\pm$ 0.11  & 0.26 $\pm$ 0.06 & 0.26 $\pm$ 0.06 \\
  2052.94 & 1                 & 1               & 1               \\
  2089.7  & 0.64  $\pm$ 0.17  & 0.68 $\pm$ 0.06 & 0.58 $\pm$ 0.09 \\
  & & & \\
  \hline
\end{tabular}
\caption[]{
	The relative intensities of the weak $\gamma$-lines from
  	$^{214}\rm{Bi}$ in the energy region from 2000 to 2100\,keV are
  	calculated with respect to the line at 2052.94\,keV. A comparison
  	between the expectations from the Table of Isotopes, the measurement
  	and the simulation is made. The discrepancy between the TOI values
  	and the measured (and simulated) values for the lines at 2010.71\,keV
  	and 2016.7 keV is due to the different position of the source.

  	($^{*}$) See note in Table 
\ref{tab2}.

  	($^{**}$) See note in Table 
\ref{tab2}.}
\label{tab3}
\end{center}
\end{table}


\newpage
\section{Conclusions}


	We presented the results obtained measuring the $^{214}\rm{Bi}$
	spectrum, with a high purity germanium detector, in the
	energy region around the the Q-value of $^{76}\rm{Ge}$ neutrinoless
	double-beta decay (2039.006\,keV). The $^{226}\rm{Ra}$ 
	source used for the measurements was positioned, 
	in a first step, directly on top of the copper cap
	of the detector and, in a second step, 15\,cm away from the copper
	cap.

	The results of the measurements show that, if the source is 
	close to the detector, the intensities of 
	the weak bismuth lines in the energy region 2000-2100\,keV are not in 
	the same ratio as reported by the Table of Isotopes. Only with a 
	simulation, which takes into account 
	the True Coincidence Summing Effect and 
	the position of the source with respect to the detector, 
	it is possible 
	to reproduce the measured intensities with good agreement.

	The analysis 
\cite{Evidence,KK02-PN,KK02-Found,KK-BigArt02} 
	of the data collected by the HEIDELBERG-MOSCOW experiment, 
	yielding a first indication for the neutrinoless
	double beta decay of $^{76}\rm{Ge}$, shows that four $^{214}\rm{Bi}$
	lines are present in the energy region from 2000 to 2080\,keV (many
	other strong lines from the same isotope are present 
	in the spectrum), due to the presence 
	of bismuth in the experimental setup, especially in the copper
	in the vicinity of the crystals.

	The present measurement shows that  
	the criticism raised in 
\cite{26people} 
	regarding the analysis of 
\cite{KK02-PN,KK02-Found,KK-BigArt02}
	of the intensities of the weak bismuth lines 
	between 2000 and 2080\,keV 
	is {\it not valid}.
	The effect of the distance between source and detector, as shown in
	this work, can affect strongly the intensities 
	of some weak lines (in the geometry investigated here by 
	a factor 1.5 for the line at 2010.71\,keV 
	and a factor 10.6 for the line at 2016.7\,keV).
	Only with a careful simulation of the experimental setup, 
	as it has been done in this work, and earlier in 
\cite{KK02-Found,replay}, 
	it is possible to correctly calculate the expected
	values for the line-intensities.


\end{document}